\documentstyle[manuscript,aps]{revtex}

\begin{document}

\title{Strong Attraction between Charged Spheres due to Metastable Ionized States}

\author{Ren\'e Messina, Christian Holm, and Kurt Kremer}

\address{Max-Planck-Institut f\"ur Polymerforschung, Ackermannweg 10,
55128, Mainz, Germany}

\date{\today{}}

\maketitle

\begin{abstract}
We report a mechanism which can lead to long range attractions between like-charged
spherical macroions, stemming from the existence of metastable ionized states.
We show that the ground state of a single highly charged colloid plus a few
excess counterions is overcharged. For the case of two highly charged macroions
in their neutralizing divalent counterion solution we demonstrate that, in the
regime of strong Coulomb coupling, the counterion clouds are very likely to
be unevenly distributed, leading to one overcharged and one undercharged macroion.
This long-living metastable configuration in turn leads to a long range Coulomb
attraction.

\end{abstract}

. 

\narrowtext

One of the great challenges in the theory of charged colloidal suspensions is
the understanding of effective attractions between like-charged macroions that
have recently been observed experimentally in confined systems {[}\ref{Kepler Fraden PRL (1994)},
\ref{Grier al - PRL (1994 - 1997) - Nature (1997)}{]}, and for which no clear
theoretical explanation is available. The usually employed mean field DLVO theory
{[}\ref{Derjaguin et al.}, \ref{Verwey et al.}{]} foresees purely repulsive
electrostatic forces between like-charged macroions. However, with divalent
counterions present, simulations (using a pair of macroions) find short range
attraction for high macroion volume fraction in aqueous systems {[}\ref{Jensen al, Physica A (1998)}{]}
or at extremely low dielectric constant a Coulomb depletion force {[}\ref{Allahyarov et al., PRL (1998)}{]}.
Recent simulations of similar systems in aqueous solutions also find attractive
forces {[}\ref{Allahyarov et al.}-\ref{Linse et al.}{]}. However all simulations
have in common, that the observed attraction occurs only for very small distances
away from the colloid surface (order of counterion size).

In this letter we investigate highly charged macroions in \textit{bulk} and
present two important new results. The first concerns the ground state of an
isolated macroion surrounded by excess counterions where it is found that the
first few overcharging counterions lower considerably the energy. As a second
finding we demonstrate that for two highly charged macroions separated by intermediate
distances thermal fluctuations are sufficient to distribute the counterions
unevenly, leading to one overcharged and one undercharged macroion. This results
in a long range effective Coulomb attraction between the macroions.

Consider one or two spherical macroions of radius \( r_{m} \) and bare charge
\( Q=-Z_{m}e \) (where \textit{e} is the elementary charge
and \( Z_{m}>0 \)  within the framework of the primitive model {[}\ref{Hill - 1960}{]}
surrounded by an implicit solvent of relative dielectric permittivity \( \epsilon _{r} \).
The small counterions with diameter \( \sigma  \) and charge +\textit{\( Z_{c}e \)}
are confined in a cubic box of length \textit{L}, and the macroion(s) are held
fixed. The colloid volume fraction \textit{\( f_{m} \)} is defined as \( N_{m}4\pi r_{m}^{3}/3L^{3} \)
(where \( N_{m} \) is the number of macroions). In the case of an isolated
macroion, it is located at the center of the box, whereas in the case of a macroion
pair, they are placed symmetrically along the axis passing by the two centers
of opposite faces.

The (MD) method employed in the present work is similar to the one used by Kremer
and Grest {[}\ref{K.}{]}. To simulate a constant temperature ensemble, the
ions are coupled to a heat bath and their motion is governed by the Langevin
equation : \( m\frac{d^{2}}{dt^{2}}\overrightarrow{r_{i}}=-\overrightarrow{\nabla }V_{tot}(\overrightarrow{r_{i}})-m\Gamma \frac{d}{dt}\overrightarrow{r_{i}}+\overrightarrow{f_{i}}(t) \),
where \textit{m} (chosen as unity) is the mass of the counterions\textit{, i}
is the \( i^{th} \) counterion, \textit{\( V_{tot} \)} is the total potential
force made up of a Coulomb term and an excluded volume term, which are both
pairwise additive, \( \Gamma  \) is the friction coefficient, and \( \overrightarrow{f_{i}} \)
a random force. These two last quantities are linked by the dissipation-fluctuation
theorem \( <\overrightarrow{f_{i}}(t)\cdot \overrightarrow{f_{j}}(t')>=6m\Gamma k_{B}T\delta _{ij}\delta (t-t^{'}) \).
For the ground state simulations the random force was set to zero.

Excluded volume interactions are introduced via a pure short range repulsive
Lennard-Jones (LJ) potential given by 
\begin{equation}
\label{math-number}
V_{LJ}(r)=4\epsilon \left[ \left( \frac{\sigma }{r-r_{0}}\right) ^{12}-\left( \frac{\sigma }{r-r_{0}}\right) ^{6}\right] +\epsilon ,\textrm{ for}\, \, r-r_{0}<r_{cut},
\end{equation}
 and 0 otherwise, where \( r_{0}=0 \) for the counterion-counterion interaction,
\( r_{0}=7\sigma  \) for the macroion-counterion interaction, \( r_{cut} \)
(= \( 2^{1/6}\sigma  \)) is the cutoff radius. This leads to \( r_{m}=7.5\sigma  \),
whereas the closest center-center distance of the small ions to the macroion
is therefore \( a=8\sigma  \). The Coulomb potential between a \( Z_{i} \)
and a \( Z_{j} \) valent ion at distance \( r \), where \textit{i} and \textit{j}
denote either macroion or counterion, is given by \( V_{coul}(r)=k_{B}T_{0}l_{B}\frac{Z_{i}Z_{j}}{r} \),
with the Bjerrum length \( l_{B}=e^{2}/4\pi \epsilon _{0}\epsilon _{r}k_{B}T_{0} \),
where \( \epsilon _{0} \) is the vacuum permittivity. To link this to experimental
units and room temperature we denote \( \epsilon = \)\( k_{B}T_{0} \) ( \( T_{0}=298 \)
K) and fix \( \sigma =3.57 \) \AA{}. We neglect hydrodynamical interactions and hydration
effects. Being interested in strong Coulomb coupling we choose, for the rest
of this paper, \( \epsilon _{r}=16 \), corresponding to \( l_{B}=10\sigma  \).

To study the possibility of overcharging a single macroion, we recall the Gillespie
rule also known as the valence-shell electron-pair repulsion (VSEPR) theory
{[}\ref{NOTE on the VESPR theory}{]}. From this one knows that the ground state
structure of two, three and four electrons disposed on a hard sphere corresponds
to simple geometrical situations, namely a line (two electrons diametrically
opposed), a triangle, and a tetrahedron, respectively. A straightforward calculation
shows that, for a central charge of +2\textit{e} , the maximally obtainable
overcharging is -2\textit{e} (i.e. 2 electrons), being independent of macroion
radius. The excess electrons gain more energy by assuming a topological favorable
configuration than by escaping to infinity, the simple reason of overcharge.
We resort to simulations to elucidate this behavior for one colloid with a high
central charge.

To quantify this phenomenon, we have considered three macroionic charge \( Z_{m} \):
50, 90 and 180 corresponding to a surface charge density of one elementary charge
per 180, 100 and 50 \AA\( ^{2} \), respectively, and fixed \textit{\( Z_{c}=2 \)}
for the rest of this letter. Then we add successively overcharging counterions
(OC). The electrostatic energy as a function of the number of OC is displayed
Fig. 1. We note that the maximal (critical) acceptance of OC (4, 6 and 8) increases
with the macroionic charge (50, 90 and 180 respectively). Furthermore for a
given number of OC, the gain in energy is always increasing with \( Z_{m} \).
Also, for a given macroionic charge, the gain in energy between two successive
overcharged states is decreasing with the number of OC. Note that at \textit{T}
= 0, the value \( \epsilon _{r} \) acts only as a prefactor. It means that
the \textit{ground state structure} is solely dictated by topological rules
(i.e., the counterions arrangement around the sphere).

The resulting curve can be very simply explained by assuming that the energy
\( \varepsilon  \) per ion on the surface of a neutralized macroion depends
linearly on the inverse distance between them, hence is proportional to \( \sqrt{N} \)
for fixed area, where \( N \) is the number of counterions on the surface.
The energy gain \( \Delta E_{1}=(N+1)\varepsilon (N+1)-N\varepsilon (N) \)
of the first OC is a pure surface correlation term. For the next OC one needs
to take into account the Coulomb repulsion \( l_{B}Z_{c}^{2}/a \), leading
to lowest order in \( 1/N \) for the energy gain of the \( n^{th} \) OC: 
\begin{equation}
\label{Eq: OC-energy}
\Delta E_{n}=n\epsilon (N)\left[ \frac{3}{2}+\frac{3n}{8N}\right] +l_{B}Z_{c}^{2}(k_{B}T_{0})\frac{(n-1)n}{2a}.
\end{equation}
 Determining \( \epsilon (N) \) from the measured value for \( \Delta E_{1} \),
we obtain a curve that matches the simulation data almost perfectly, compare
Fig.1.

An energy per ion, which scales like \( \sqrt{N} \), has been found for an
ionic Wigner crystal (WC) on a planar surface where each ion interacts with an oppositely
charged background charge which is smeared out over its Wigner-Seitz cell. This
energy per ion is given by \( \varepsilon (c)/k_{B}T_{0}=-\alpha c^{1/2}l_{B}Z_{c}^{2} \),
with \( \alpha =1.96 \), and \( c \) is the two-dimensional concentration
of the crystallized counterions of valence \( Z_{c} \) \cite{maradudin77a}.
This Ansatz has been tried recently to explain strong ionic correlations observed
in various soft matter system \cite{shklovskii99a,shklovskii99b}. In our simulation
we find for \( \Delta E_{1}/(k_{B}T_{0}) \) -18.0, -24.4, and -35.3 for \( Z_{m}=50 \),
90, and 180, respectively, whereas the Wigner crystal scenario predicts -21.0,
-28.0, and -39.5, which is off by a decreasing rate of \( 17-12\% \). This
might be due to the assumption of a homogeneous background charge, and the assumption
of a planar geometry, neither of which are fully fulfilled, however the error
gets smaller for higher values of \( Z_{m} \).

Using the Wigner crystal ionic energy and eq. (\ref{Eq: OC-energy}), the maximally
obtainable number \( n^{*}_{max} \) of OC counterions is readily found to be
\begin{equation}
\label{Q*}
n^{*}_{max}=\frac{1}{2}+\frac{9\alpha ^{2}}{32\pi }+\frac{3\alpha }{4\sqrt{\pi }}\sqrt{N}+\left[ \frac{3\alpha }{16\sqrt{\pi }}+\frac{27\alpha ^{3}}{256\pi ^{3/2}}\right] \frac{1}{\sqrt{N}}+{\mathcal{O}}(1/N).
\end{equation}
 This value depends only on the number of counterions \( N \). It originates
from the topological arrangement of the ions around a central charge, and is
independent of Bjerrum length or radius of the macroion. For large \textit{Q}
it reduces to the form \( Q^{*}_{max}/e\approx \frac{3\alpha }{4\sqrt{\pi }}*\sqrt{Z_{m}Z_{c}} \)
which was derived in Ref. \cite{shklovskii99b} in a more elaborate fashion.

To obtain the interaction potential profile, we added one counterion coming
from infinity towards a macroion of bare charge \( Z_{m}=180 \) and computed
the global electrostatic energy of the system, see Fig. 2. The first OC starts
to gain correlational energy at distance \( r\approx 12\sigma  \) from the
center of the colloid, which is about \( 4\sigma  \) from the surface. This
fits only roughly with the distance \( Z_{c}^{2}l_{B}/4 \) predicted from WC
theory \cite{shklovskii99a,shklovskii99b}, and is more of the order \( c^{-1/2} \).
With adding more excess counterions the Coulomb barrier increases, and for the
ninth OC it exceeds the gain in correlational energy, when being on the macroion
surface. Thus the configuration becomes metastable. The curve for the \textit{first}
OC can be nicely fitted with an exponential fit of the form \( E_{1}(r)/k_{B}T_{0}=-35.3\exp {[-7.1(r-a)/a]} \).
For the \( n^{th} \) OC simply the appropriate Coulomb monopole contribution
\( 4l_{B}(n-1)/r \) needs to be added, see Fig.2. This exponential dependence
is not predicted by the WC theory, where a 1/r dependence should be seen due
to the interaction of the removed ion with its correlation hole.

Next, we consider two spherical like charged macroions at a colloidal volume
fraction \( f_{m}=7\cdot 10^{-3} \) at \textit{room temperature} \textit{T\( _{0} \)},
at fixed center-center separation \textit{R}, in presence of their divalent
counterions (ensuring global charge neutrality). Initially the counterions are
randomly generated. Figure 3 shows two macroions surrounded by their quasi-two-dimensional
counterions layer. The striking peculiarity in this configuration is that it
corresponds to an overcharged and an undercharged sphere. There is one counterion
more on the left sphere and one less on the right sphere compared to the bare
colloid charge. Such a configuration is referred as \textit{ionized state}.
In a total of 10 typical runs, we observe this phenomenon 5 times. We have also
carefully checked against a situation with periodic boundary conditions, yielding
identical results. However it is clear that such a state is in ``pseudo-equilibrium''
because it is not the lowest energy state.

To estimate the energy barrier, electrostatic energy profiles at \textit{zero
temperature} were computed, where we move one counterion from the overcharged
macroion to the undercharged, restoring the neutral state (see Fig. 4). We have
checked that the path leading to the lowest barrier of such a process corresponds
to the line joining the two macroions centers. One clearly observes a barrier,
which increases linearly with the charge \textit{\( Z_{m} \)}. The ground state
corresponds as expected to the neutral state. The overcharged state is only
slightly higher in energy, the difference being approximately the monopole contribution
\( E/k_{B}T_{0}=l_{B}(4/8-4/12)\approx 1.67 \). The physical origin of this
barrier can be understood from the single macroion case where we showed that
a counterion gains high correlational energy near the surface. This gain is
roughly equal for both macroion surfaces and decreases rapidly with increasing
distance from the surfaces, leading to the energy barrier with its maximum near
the midpoint. For the single macroion case we showed that the correlational
energy gain scales with \( \sqrt{Z_{m}} \), whereas here we observe a linear
behavior of the barrier height with \( Z_{m} \). We attribute this effect to
additional ionic correlations since both macroions are close enough for their
surface ions to interact strongly. For large separations we find again that
the barrier height increases with \( \sqrt{Z_{m}} \), as expected. This \( Z_{m} \)
dependence of the barrier also shows that at room temperature such ionized states
only can occur for large \( Z_{m} \). In our case only for \textit{\( Z_{m}=180 \)},
the ionized state was stable for all accessible computation times. Unfortunately,
it is not possible to get a satisfactory accuracy of the energy jumps at non-zero
temperatures. Nevertheless, since we are interested in the strong Coulomb coupling
regime, which is energy dominated, the zero temperature analysis is sufficient
to capture the essential physics.

Results concerning the effective forces at \textit{zero temperature} between
the two macroions are now investigated which expression is given by 
\begin{equation}
\label{math-number}
F_{eff}(R)=F_{mm}(R)+F_{LJ}+F_{mc}\, ,
\end{equation}
 where \( F_{mm}(R) \) is the direct Coulomb force between macroions, \( F_{LJ} \)
is the excluded volume force between a given macroion and its surrounding counterions
and \( F_{mc} \) is the Coulomb force between a given macroion and all the
counterions. Because of symmetry, we focus on one macroion. To understand the
extra-attraction effect of these ionized-like states, we consider three cases:
(i) \( F_{ion}=F_{eff} \) in the ionized state (ii) \( F_{neut}=F_{eff} \)
in the neutral case (iii) \( F_{mono}=F_{eff} \) simply from the effective
monopole contribution. Our results are displayed in Fig. 5 for \( Z_{m}=180 \),
where the ionized state was also observed at room temperature. The non-compensated
case leads to a very important extra attraction. This becomes drastic for the
charge asymmetry of \( \pm  \) 2 counterions at short separation \( R/a\approx 2.5 \),
a situation which was also observed in our simulation at room temperature {[}\ref{For the +/- 2 counterions case}{]}.
In contrast to previous studies {[}\ref{Jensen al, Physica A (1998)}, \ref{Allahyarov et al., PRL (1998)}{]},
these attractions are long range. For a sufficiently large macroion separation
(from 3.5) the effective force approaches in good approximation the monopole
contribution.

In summary, we have shown that a sufficiently charged colloid can in principle
be highly overcharged due to correlation effects of the counterions, and that
this effect is quantitatively well described by a Wigner crystal, i.e. eqs.
(\ref{Eq: OC-energy}) and (\ref{Q*}). In the strong Coulomb coupling regime,
this energy gain can be of the order of many \( k_{B}T_{0} \).

Furthermore, due to this energetically favorable overcharged state it was found
that for two like-charged macroions, an initially randomly placed counterion
cloud of their neutralizing divalent counterions may not be equally distributed
after relaxation, leading to two macroions of opposite net charges. The resulting
configuration is metastable, however separated by an energy barrier of several
\( k_{B}T_{0} \) when the bare charge is sufficiently large. Such configuration
possess a natural strong long range attraction.

We acknowledge useful discussions with H. Löwen and M. Deserno. This work is
supported by \textit{Laboratoires Europ\'eens Associ\'es} (LEA) and a computer time
grant hkf06 from NIC J\"{u}lich.


\listoffigures{}

FIG. 1: Electrostatic energy (in units of \( k_{B}T_{0} \)) for \textit{zero
temperature} configurations of a single charged macroion of radius \textbf{\( r_{m}=7.5\sigma  \)}
as a function of the number of overcharging counterions for three different
bare charges \( Q \) (in units of \textit{e} ). The neutral case was chosen
as the potential energy origin, and the curves were produced using the theory
of eq.(2), compare text.

FIG. 2: Electrostatic energy (in units of \( k_{B}T_{0} \)) of a divalent counterion
as function of distance from the center of a macroion with radius \( r_{m}=7.5\sigma  \)
and charge \( Q=-180 \) (in units of \textit{e} ). The energy is normalized
to zero at distance infinity. Data and fits are shown for the first, the second,
the eighth and ninth overcharging (OC) counterion.

FIG. 3 Snapshot of a ``pseudo-equilibrium'' configuration at room temperature
\( T_{0} \) where the counterion-layers do not exactly compensate the macroions
charge. Here the deficiency charge is \( \pm  \)1 counterion (or \( \pm  \)
2\textit{e} as indicated above the macroions) and \textit{R}/\textit{a} = 3.6\textit{.}

FIG. 4 Total electrostatic energy (in units of \( k_{B}T_{0} \)) of the system,
for \textit{zero temperature} configurations, of two macroions at a center-center
separation of \textit{R}/\textit{a} = 2.4 as a function of one displaced counterion
distance from the left macroion for three typical values \( Q \) (in units
of \textit{e} ). The exact neutral state was chosen as the potential energy
origin. The lines are a guide to the eye. The insert indicates the path (dotted
line) of the moved counterion. The ending arrows of the arc indicate the start
position (left sphere) and final position (right sphere) of the moved counterion.

FIG. 5 Reduced effective force between the two spherical macroions at \textit{zero
temperature} for \( Z_{m}=180 \) as a function of distance from the center.
The different forces are explained in the text. The lines are a guide to the
eye.

\end{document}